# COSMOLOGICAL CONSTANT AND POLYMER PHYSICS

P. R. Silva – Departamento de Física – ICEX – Universidade Federal de Minas Gerais
C. P. – 702 - 30123-970 – Belo Horizonte – MG – Brazil
e-mail1: prsilvafis@terra.com.br - e-mail2: prsilva@fisica.ufmg.br

ABSTRACT – It is traced out a parallel between the cosmological constant problem and the polymer physics. The time evolution of the universe world line is compared with the growing of a polymer chain. An equivalent Flory free energy and a modification of it are used as a means to evaluate two versions for the "radius of gyration" of the universe. It is proposed a link between this radius of gyration and the cosmological constant parameter.

INTRODUCTION

  The small value of the observed cosmological constant [1] can not be explained by the value of the vacuum energy density as predicted by the quantum field theory of the standard-model degrees of freedom [2]. The smallness of the cosmological constant has been considered as one of the central problems of the theoretical physics [3]. Besides this, as was pointed out by Ke and Li [4], recently there has been considerable interest in explaining the observed dark energy [5,6,7] by the holographic dark energy model [8,9]. In their holographic dark energy model Cohen, Kaplan and Nelson [8], found that the cosmological constant is determined by a relation which involves the radius of the observed universe and the Planck mass. This came from a conection between ultra violet and infra red cutoffs, such that an effective field theory should be a good description of nature.
  Recently Hsu and Zee [10] made a proposal of an effective action for our universe. This action is a sum of a term linear in the cosmological constant multiplied by a four-volume, where the length scale is that characteristic of the observable universe, plus a term which depends inversely on the cosmological constant multiplied by the fouth power of the Planck mass. The second term of the action of Hsu and Zee [10] was explained by Ke and Li [4] as due to quantum corrections (related to a kind of Casimir energy). Minimizing their effective action, Hsu and Lee obtained a relation for the cosmological constant which reproduces previous results found by Cohen and collaborators [8].
  Meanwhile, as was pointed out by Hsu a nd Zee [10], Padmanabhan [11] has proposed that gravity only couples to the fluctuations of the vacuum energy rather than the vacuum energy itself. Working in this way, Padmanabhan obtained a relation for the cosmological constant which is similar to those found in [8] and [10].
   On the other hand, the simplest type of polymer are represented by linear chains of monomers [12]. According to Raposo et al [12], in the lattice model of polymers, a dilute macromolecule in a good solvent, is represented by a random self-avoiding walk (SARW) on a regular lattice.
  Inspired in these ideas of the polymer physics, we devised the possibility of describing the time evolution of the univese as a grow of a polymer chain. As polymers chains grow by the addition of monomers units of fixed length, we are led to think these small units as fixing the high violet cut off of the model. Meanwhile we could consider that some features of the string physics [16] may be relevant to the present work due to the one dimensional character of both of these mathematical objects.

A brilliant scheme for computing the radius of gyration of a polymer (the end-to-end distance of a SARW) was devised by Flory long time ago [13], for the three dimensional case. Later on, Fisher [14] extended the Flory results to other dimensions. In the following, we use the arguments presented in de Gennes book [15], in order to introduce the Flory free energy.

The starting point is a chain, with a certain unknown radius R and a internal monomer concentration

$$c_{int} = N/R^d, \qquad (1)$$

where d stands for the fact that the calculation is peformed on a d-dimensional lattice. There is a certain repulsive energy in the chain due to monomer-monomer interactions. This repulsive energy per unit volume is proportional to $c^2$, the number of pairs present. We write

$$f_{rep} = 1/2 \; t \; v \; c^2, \qquad (2)$$

where t is the temperature and v is the excluded volume parameter (see [15]). After integrating in a volume $R^d$, we find that the overall repulsive energy scales as

$$F_{rep} = f_{rep} \; R^d = t \; v \; N^2/R^d. \qquad (3)$$

However a linear ideal chain in any dimension has a a gaussian distribution for R, due to its random walk character, given by (see [15])

$$p(R) = p_0 \exp[-R^2/(Na^2)], \qquad (4)$$

where a is the lattice parameter. Relation (4) leads to an entropic contribution for the free energy the so-called elastic energy

$$F_{el} = t \; R^2/(Na^2). \qquad (5)$$

Flory free energy is given by the sum of terms (3) and (5), namely

$$F/t = v \; N^2/R^d + R^2/(Na^2). \qquad (6)$$

Minimizing (6) relative to R, we obtain Flory's Radius $R_F$ which scales as

$$R_F \sim N^\upsilon, \qquad (7)$$

where $\upsilon$ is given by

$$\upsilon = 3/(d+2). \qquad (8)$$

For d = 4, relation (8) gives $\upsilon$ = ½, the ideal chain exponent (pure random walk).

THE FIRST MODEL

Inspired in the Flory's free enegy model of an ideal polymer, we propose an alternative model to treat the cosmological constant problem. The basic hypothesis of the model are:
i) The universe world line will be represented by a segmented line, as a lattice model of a polymer (SARW).
ii) The lattice spacing will be taking as the Planck length $L_P$, which naturally sets up the ultra violet cutoff of the problem.
iii) The walk can not make self-intersections, due to the excluded volume constraint.
iv) The extension of the chain will be take as $L = NL_P$, where N is the number of steps, and L will be interpreted as the radius of of the observable universe, fixing the infra red cutoff of the problem. We observe that the present treatment considers that both time (N) and space ($L_P$) evolve through discrete accretion (steps).
v) The radius of gyration or end to end distanc $R_\Lambda$ of the SARW, will be interpreted as the Compton length associated to the cosmological constant or dark energy.

Thus we write the Flory energy of the universe $F_u$, as

$$F_u / t = N^2 L_P^d / R^d + R^2 / (NL_P^2), \qquad (9)$$

where t is un unspecified temperature.
Setting $L = NL_P$ and minimizing (9) relative to R, we obtain for the cosmological radius $R_\Lambda$, the relation

$$R_\Lambda(d) = L^{3/(2+d)} L_P^{(d-1)/(2+d)}. \qquad (10)$$

In the above relations, d is the space-time dimension. The d=4 dimension deserves special attention. We have:

$$R_\Lambda(d=4) = L^{1/2} L_P^{1/2}. \qquad (11)$$

If we define as Hsu and Zee [10] a mass scale $M_u = L^{-1}$ and $M_\Lambda = R_\Lambda^{-1} = \Lambda^{1/4}$, we have:

$$M_\Lambda = (M_u M_P)^{1/2}, \qquad (12)$$

which reproduces relation (4) of Hsu and Zee [10] paper. Also if we take $L_P = 10^{-35}$ m and $L = 10^{26}$ m, we get $N = 10^{61}$ and

$$R_\Lambda(d=4) = \sqrt{10} \times 10^{-5} \text{ m}. \qquad (13)$$

$R_\Lambda(4)$ obtained in (13) perhaps could be interpreted as the radius of gyration ( or coherence length ) of the universe. It is interesting to note that relation (10) also can be put in the form

$$R_\Lambda = N^{3/(2+d)} L_P, \qquad (14)$$

where $3/(2+d)$ is the Flory's exponent.
If we identify $R_\Lambda$ obtained in (13) with the Compton wavelength of a particle of mass $M_\Lambda$, the mass-energy of this particle will be approximately $5 \times 10^{-3}$ eV, and this value is comparable in order of magnitude with the fourth root of the observed dark energy density [5,6].

THE SECOND MODEL

In the second model we will retain some elements of the Flory's theory, but introducing besides this some important modifications. The new basic hypothesis of this second model are
  a) We suppose that the chain which represents the universe world line is embedded in a n-dimensional space-like volume. In this way, space and time are no more treated at equal footing.
  b) The entropy of the model is evaluated by counting the number of "monomers units" of length $L_P$, necessary to fill up the line of length R.
  With these ideas in mind, we write the new free energy as

$$F_n/t = N^2 L_P^n/R^n + R/L_P. \tag{15}$$

In equation (15) the first term takes in account the feature a) of the model, while the feature b) is accounted by the second term.
  The entropic contribution for the free energy (15) (see feature b)) may be justified through the following reasoning. The number of steps of size $L_P$ needed to fill up a chain of length L is given by

$$N = L / L_P. \tag{16}$$

Indeed, Kalyana Rama [16] in a paper dealing with the size of black holes through polymer scaling found an entropy given by

$$S_{string} = a\, M, \tag{17}$$

where a is a string scale and M is its mass. Making the requirement that the string scale be equals to $L_P$, we can rewrite (17) as

$$S_{string} = M / M_P = L / L_P. \tag{18}$$

In (18) we have considered that the mass density of the chain is constant up to the string scale $L_P$.
Inspired in Padmanabhan idea [11], we propose that in the case of the cosmological constant are the fluctuations in the number of steps which contribute for the entropy of the problem and write

$$S_{cosmological} = ( L / L_P )^{1/2}. \tag{19}$$

Taking in account (11) we get

$$S_{cosmological} = ( R / L_P ), \tag{20}$$

which leads to the entropic contribution to (15).
Now taking the extremum of (15) relative to R, we obtain

$$R^{n+1} = L^2 L_P^{n-1}, \tag{21}$$

where $L = NL_P$.

Alternatively by setting $R^{n+1} = (\Lambda^{n+1})^{-1}$, we get

$$\Lambda^{n+1} = M_u^2 M_P^{n-1}. \tag{22}$$

In the case of three spatial dimensions (n=3), relation (22) reproduces previous results both of Hsu and Zee [10] and Cohen, Kaplan and Nelson [8]. Our main result, equation (22) which generalizes to n-spatial dimensions, can be compared with the work of Ke and Li [4]. They used Cardy-Verlinde [17] and Gibbons-Hawking entropy to obtain the Casimir energy to the problem and after, minimizing the total energy, they got the dark energy formula. Thus relation (22) of this paper must be compared with equation (7) of Ke and Li [4], for the holographic dark energy in n+1 dimensions. To compare the two results, we must consider that $\Lambda$ and R of the work of Ke and Li corresponds respectively to $\Lambda^{n+1}$ and L of the present work Also, as noted by Ke and Li [4], equation (22) is consistent with a (n+1)-dimensional de Sitter space, if we identify in the Friedman equation in n+1 dimensions, the matter density with the cosmological constant $\Lambda^{n+1}$ and the Hubble constant H with $L^{-1}$ (the inverse of the cosmic horizon size for a de Sitter space).

CONCLUDING REMARKS

In closing this paper, we judge worth to add two pieces of reasoning that we suppose will be relevant to best clarify the present work.
First we discuss equation (9), namely the Flory energy of the universe in the case d=4. By considering $R^4 = 1/\Lambda$, and $L = N L_P$, we can write

$$F_u/t(d=4) = L^2 L_P^2 \Lambda + [(L L_P) \Lambda^{1/2}]^{-1}. \tag{23}$$

Equation (23) keeps some resemblance with the action proposed by Hsu and Zee [10], with a term proportionl to $\Lambda$ but integrated in a four-volume which linear dimension is the geometric average between L and $L_P$, plus a term which goes as $\Lambda^{-1/2}$ and therefore does not behaves smoothly as $\Lambda$ goes to zero through positive values.

As a second remark, perhaps the simplest way of taking in account the contribution of the cosmological constant to the evolution of the universe can be found in a paper by Guth and Steinhardt [18], please see also [19]. Let us suppose a sphere of mass M homogeneously distributed and radius r and a test mass m localized at its surface. We consider that the test mass is influenced by a potential energy which is a sum of an attractive gravitational term due to the sphere of mass M plus a repulsive term due to the cosmological constant (or dark energy) and write

$$E_{pot} = - GMm/r - 1/6\, \Lambda^{1/2} m\, c^2\, r^2. \tag{24}$$

In order to proceed with the analysis of the problem, let us take the total energy as

$$E_{total} = \tfrac{1}{2} m v^2 + E_{pot}, \qquad (25)$$

where both Hubble's law (v = Hr), and the cosmological constant term $\Lambda$, are inserted in (25) by the hand. This procedure leads to the Friedmann equation for the ordinary matter and dark energy.

Finally, we would like to stress that the calculations performed in this paper, neglects various constants parameters which would appear in the formulas for the free energies, assuming that they would be of the order of the unity.